\documentclass[3p,authoryear]{elsarticle}
\usepackage{epstopdf}
\usepackage{subfigure,graphicx}
\usepackage{float}    
\usepackage{amsmath, amsfonts, amssymb,mathrsfs}
\usepackage{amsthm}
\usepackage{epsf}
\usepackage{amsfonts}
\usepackage{stmaryrd}
\usepackage{amssymb}
\usepackage{leftidx}
\usepackage{color}
\usepackage{mathtools}
\usepackage{placeins}
\usepackage{booktabs}
\usepackage{enumitem}
\usepackage{caption}
\usepackage{multirow}
\usepackage{stackengine}
\usepackage[utf8]{inputenc}
\usepackage[english]{babel}
\usepackage{color}

\theoremstyle{plain}

\def\bfe{{\bf e}}

\def\bfs{{\bf s}}

\def\bfv{{\bf v}}

\def\bfy{{\bf y}}

\def\bfI{{\bf I}}

\def\bfN{{\bf N}}

\def\bfS{{\bf S}}
\def\bfT{{\bf T}}
\def\bfX{{\bf X}}
\def\bfF{{\bf F}}

\def\obfF{\overline{\bfF}}

\def\ol{\overline{\lambda}}

\long\def\symbolfootnote[#1]#2{\begingroup%
\def\thefootnote{\fnsymbol{footnote}}\footnote[#1]{#2}\endgroup}

\makeatletter
\renewcommand\@biblabel[1]{}

\makeatother

\begin{document}
\begin{frontmatter}

\title{The nonlinear elastic deformation of liquid inclusions \\ embedded in elastomers\vspace{0.1cm}}

\author{Oluwadara Moronkeji}

\author{Fabio Sozio}

\author{Kamalendu Ghosh}

\author{Amira Meddeb}

\author{Amirhossein Farahani}

\author{Zoubeida Ounaies}

\author{Ioannis Chasiotis}

\author[Illinois]{Oscar Lopez-Pamies}
\ead{pamies@illinois.edu}

\address{Department of Aerospace Engineering, The Grainger College of Engineering, \\ University of Illinois, Urbana--Champaign, IL 61801, USA \vspace{0.1cm}}

\address{Department of Civil and Environmental Engineering, The Grainger College of Engineering, \\ University of Illinois, Urbana--Champaign, IL 61801, USA  \vspace{0.1cm}}

\address{Department of Mechanical Engineering, Pennsylvania State University, University Park, PA 16802, USA \vspace{0.1cm}}

\address{Department of Materials Science and Engineering, Pennsylvania State University, University Park, PA 16802, USA \vspace{0.1cm}}

\vspace{-0.1cm}

\begin{abstract}

Elastomers filled with liquid inclusions --- as opposed to conventional solid fillers --- are a recent trend in the soft matter community because of their unique range of mechanical and physical properties. Such properties stem, in part, from the very large deformations that the underlying liquid inclusions are capable of undergoing. With the objective of advancing the understanding of the mechanics of this emerging class of materials, this paper presents a combined experimental/theoretical study of the nonlinear elastic deformation of initially spherical liquid inclusions embedded in elastomers that are subjected to quasistatic mechanical loads. The focus is on two fundamental problems, both within the limit regime when elasto-capillarity effects are negligible: ($i$) the problem of an isolated inclusion and ($ii$) that of a pair of closely interacting inclusions. Experimentally, specimens made of a polydimethylsiloxane (PDMS) elastomer filled with either isolated or pairs of initially spherical liquid glycerol inclusions are subjected to  uniaxial tension. For the specimens with pairs of inclusions, three orientations of the two inclusions with respect to the direction of the applied macroscopic tensile load are considered, $0^\circ$, $45^\circ$, and $90^\circ$. The liquid glycerol is stained with a fluorescent dye that permits to measure the local deformation of the inclusions \emph{in situ} via confocal laser scanning fluorescent microscopy. Theoretically, a recently developed framework --- wherein the elastomer is considered to be a nonlinear elastic solid, the liquid comprising the inclusions is considered to be a nonlinear elastic fluid, and the interfaces separating the elastomer from the liquid inclusions can feature their own nonlinear elastic behavior (e.g., surface tension) --- is utilized to carry out full-field simulations of the experiments. \emph{Inter alia}, the results show that the deformation of liquid inclusions is significantly non-uniform and strongly influenced by the presence of other liquid inclusions around them. Interestingly, they also show that the large compressive stretches that localize at the poles of the inclusions may result in the development of creases.

\keyword{Elastomers; Fillers; Elasto-capillarity; Finite deformation; Homogenization}
\endkeyword

\end{abstract}

\end{frontmatter}

\vspace{-0.1cm}

\section{Introduction}\label{Sec:Intro}

By now, it is well understood that the use of \emph{liquid} inclusions --- as opposed to conventional \emph{solid} inclusions --- as fillers in elastomers may result in materials with unique properties because of two main reasons. The first is that the addition of liquid inclusions to elastomers increases the overall deformability \citep{LP14,Bartlettetal2017,LDLP17,Yunetal19}. This is in contrast to the addition of conventional fillers, which, being typically made of stiff solids, decreases deformability. The second main reason behind the fascinating properties of elastomers filled with liquid inclusions is that the behavior of the interfaces separating a solid elastomer from embedded liquid inclusions feature their own mechanical/physical behavior, one that, while negligible when the inclusions are ``large'', may dominate the macroscopic properties of the material when the inclusions are sufficiently ``small'' \citep{Syleetal15,Andreottietal16,Bicoetal18,GLP22}.

In contrast to the well-established qualitative understanding outlined above, the quantitative understanding of the mechanics and physics of elastomers filled with liquid inclusions is yet to be fully developed. Towards that end, a string of recent works by \cite{GLP22}, \cite{GLLP23a,GLLP23b}, and \cite{CDFLPM24} have introduced a series of theoretical results that include the derivation of the governing equations that describe the mechanical response of nonlinear elastic solids filled with initially spherical inclusions made of pressurized nonlinear elastic fluids when the solid/fluid interfaces are nonlinear elastic and possess an initial surface tension, as well as the homogenization limit of these equations. This paper is the next installment of this series. 

Precisely, this paper presents a combined experimental/theoretical study of the nonlinear elastic deformation of initially spherical liquid inclusions embedded in elastomers that are subjected to quasistatic mechanical loads. The focus is on two fundamental problems: ($i$) the problem of an isolated liquid inclusion and ($ii$) the problem of a pair of closely interacting inclusions. Both of these problems are considered within the limit regime when elasto-capillarity effects are negligible. Experimentally, specimens made of a polydimethylsiloxane (PDMS) elastomer filled with either isolated or pairs of initially spherical liquid glycerol inclusions are subjected to uniaxial tension. The initial shear modulus of the elastomer is about $\mu=45.5$ kPa, the initial surface tension at the interfaces between the elastomer and the liquid glycerol is about $\gamma=0.014$ N/m, and the initial radii of the inclusions range from $A=35$ $\mu$m to $264$ $\mu$m so that the initial  elasto-capillary number ranges from $eCa:=\gamma/2\mu A=5.8\times 10^{-4}$ to $4.4\times 10^{-3}$. These values are small enough that elasto-capillarity effects can indeed be neglected \citep{GLLP23b}. As a key aspect, the liquid glycerol is stained with a fluorescent dye that permits to measure the local deformation of the inclusions \emph{in situ} via confocal laser scanning fluorescent microscopy. Theoretically, the framework introduced by \cite{GLP22} is utilized to carry out full-field simulations of the experiments. 

We begin in Section \ref{Sec:Materials} by describing the materials, specimen preparation, and experimental techniques used in this study. The theoretical description of the experiments is laid out in Section \ref{Sec:Theory}. Section \ref{Sec:Results} is devoted to presenting and discussing the experimental results alongside their full-field simulations. We conclude in Section \ref{Sec:Final comments} by recording a number of final comments.

\subsection{A summary of previous results for dilute suspensions of spherical inclusions in liquids and solids} 

Before proceeding with the presentation of the proposed study \emph{per se}, it is instructive to provide a summary of the long and rich history of the mechanics of dilute suspensions of initially spherical inclusions, in both liquids and solids.  

For dilute suspensions of initially spherical inclusions in \emph{liquids}, the first main result in the literature is famously due to \cite{Einstein06,Einstein11}, who worked out  the exact solution for the Stokes flow of an incompressible Newtonian fluid embedding an isolated rigid inclusion. About three decades later, \cite{Taylor32} worked out the analogous solution for the case of an isolated inclusion made of another incompressible Newtonian fluid taking into account the presence of a large surface tension at the interface between the two fluids, which kept the inclusion from deforming. Five decades later, \cite{BG72} considered the problem of a pair of closely interacting inclusions\footnote{For the mathematical analysis of this problem and further historical notes, the interested reader is referred to the recent monograph by \cite{Gloria2023} and references therein.}, made either of a solid or of another incompressible Newtonian fluid, taking into account again the presence of a large surface tension so as to prevent the deformation of the inclusions. In the footsteps of these three classical works, significant efforts have been and continue to be devoted to generate solutions, within the setting of Stokes flow as well as of flows with high Reynolds numbers, for the more challenging problem of isolated inclusions and pairs of inclusions that are initially spherical in shape but that eventually deform and even break up as a result of the flow that they are subjected to; see, e.g., \cite{Stone94}. 

On the other hand, for dilute suspensions of spherical inclusions in \emph{solids}, exploiting the mathematical equivalency between Stokes flow and the equations of linear elastostatics
at a fixed instant in time, \cite{Smallwood44} transcribed the solution of \cite{Einstein06,Einstein11} to that of an isolated rigid inclusion in an incompressible linear elastic solid. Subsequently, \cite{Eshelby57} famously worked out the solution for the more general case of an isolated inclusion made of an isotropic linear elastic solid embedded in another isotropic linear elastic solid. About two decades afterwards, still within the setting of small-deformation linear elastostatics, \cite{Acrivos78} derived the solution for a pair of interacting inclusions. The first results in the general setting of finite deformations were put forth more recently, by \cite{LPGD13a} and \cite{LLP17a}, who worked out the solutions for isolated rigid and liquid inclusions in an incompressible Neo-Hookean solid. 

All the above four results for suspensions of spherical inclusions in solids assumed the absence of surface tension at the solid-matrix/inclusion interface. In the asymptotic setting of small deformations, the case of an isolated liquid inclusion accounting for the presence of surface tension at the interface between the linear elastic solid and the liquid inclusion was considered (as an Eulerian problem) about a decade ago  by \cite{Syleetal15b}. As part of the same effort, \cite{Syleetal15} reported experimental results for the deformation of ionic-liquid inclusions of several initial radii $A\in[2,38]$ $\mu$m, and hence several initial elasto-capillary numbers $eCa\in[0.0836, 1.5882]$, embedded at dilute concentrations in a soft silicone elastomer subjected to finite deformations of different non-uniaxial triaxialities. On the theoretical front in the general setting of finite deformations, just a few years ago, \cite{GLP22} worked out the solution (making use of a Lagrangian formulation) for an isolated liquid inclusion in an incompressible Neo-Hookean solid, wherein the solid-matrix/liquid-inclusion interface features a constant surface tension. 

In this context, the experimental and theoretical results presented in this work for isolated and pairs of liquid inclusions embedded in nonlinear elastic solids undergoing finite deformations can be viewed as natural additions to both the fluids and solids literatures on dilute suspensions of spherical inclusions.

\section{Materials, specimens, and experimental methods}\label{Sec:Materials}

\subsection{Materials and fabrication methods}

In this work, the elastomer used in all of the fabricated specimens is the popular PDMS Sylgard 184 supplied by Dow Corning as a two-part kit, which comprises a base elastomer and a curing
agent. Liquid glycerol was used for the inclusions. 

The choice of PDMS Sylgard 184 was motivated by several of its attributes. It is a class of elastomers that exhibit almost purely elastic behavior with very little viscous dissipation and Mullins effect, provides facile access to a wide range of cross-link densities, and is transparent and thus amenable to optical microscopy through its volume. The choice of glycerol was motivated for its chemical simplicity, transparency, and ability to be easily stained. 

Two methods were employed to fabricate PDMS filled with liquid glycerol inclusions. The first method allowed to fabricate films with a dilute volume fraction of randomly distributed inclusions of initial radii in the range $A\in[35,100]$ $\mu$m. The second method allowed to fabricate films with a precise placement of the inclusions in the PDMS, at the expense of their initial radii being larger, in the order of $A=250$ $\mu$m.

\begin{figure}[t!]
\centering
\centering\includegraphics[width=0.55\linewidth]{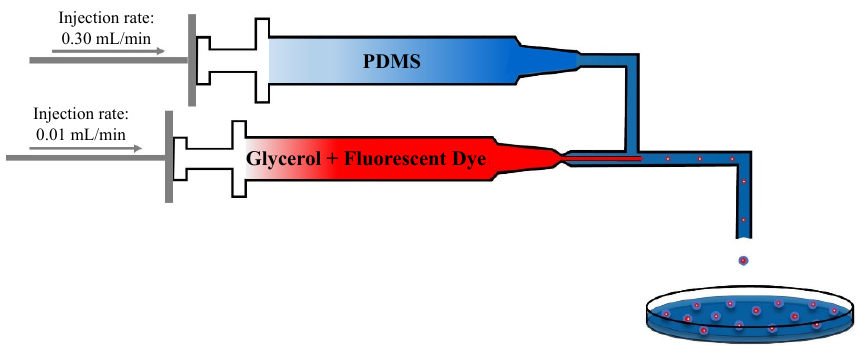}
\caption{\small Schematic of the coaxial microfluidic device (CMD) used to fabricate films of PDMS filled with liquid glycerol inclusions stained with a fluorescent dye, Sulfo-Cyanine3 (Cya3) amine.}\label{Fig1}
\end{figure}

Precisely, as illustrated schematically in Fig. \ref{Fig1}, the first method consisted in introducing stained glycerol droplets into PDMS with the aid of a coaxial microfluidic device (CMD) featuring two reservoirs. The stained glycerol was made by first mixing glycerol with a small amount (0.004$\%$ in weight fraction) of Sulfo-Cyanine3 (Cya3) amine fluorescent dye. Methanol was subsequently added to the mix at a volume ratio of 5:1 with respect to glycerol (i.e., 1 part methanol for 5 parts glycerol) to reduce viscosity, followed by mixing in a non-contact planetary (THINKY) mixer at 500 rpm for 30 s, with an additional 30 s at 1,000 rpm under vacuum. After degassing, the solution was transferred to one of the two reservoirs in the CMD; see Fig. \ref{Fig1}. As for the solution fed to the other reservoir in the CMD, it comprised a mix of the base elastomer with the curing agent, the solvent $n$-heptane, and a surfactant at weight ratios 33:1, 10:1, and 100:1 with respect to the base elastomer. The entire mixture underwent thorough mixing and degassing in a THINKY mixer before being transferred to the CMD. To fabricate the samples, pure PDMS solution was first poured into a Teflon petri dish and partially cured at 100 $^\circ$C for 30 minutes. Then both the PDMS and the glycerol solutions were poured at the rates of 0.30 ml/min and  0.01 ml/min, respectively. The resulting sample was fully cured at 100 $^\circ$C for 4 hours.

In the second method, droplets of stained glycerol were manually injected into PDMS via a syringe. Specifically, the base elastomer was first mixed with a surfactant (1.2$\%$ in weight fraction) in a THINKY mixer, followed by the addition of the curing agent at the weight ratio 33:1 with respect to the base elastomer. To fabricate the samples, much like in the first method, a first layer of PDMS was cast and precured at 100 $^\circ$C for 30 min. The remainder of the PDMS mixture was stored in a refrigerator at 4 $^\circ$C.  Before reusing the refrigerated PDMS mixture to cast the next layers, it was mixed in a THINKY mixer. Meanwhile, the fluorescent dye Cya3 was added to glycerol at a 0.004$\%$ in weight fraction, and the resulting solution was agitated in a bath sonicator for 10 min. The second layer of PDMS was cast over, and Cya3-glycerol droplets were manually injected into it using a G36 needle (World Precision Instruments syringe). The second layer was then cured at 100 $^\circ$C for 30 min. Finally, a third layer was cast upon the second layer, and the sample was oven cured at 100 $^\circ$C for an additional 4 hours.

The two fabrication methods described above yield films, of 20 cm in diameter and about 1 mm in thickness, comprised of PDMS filled with glycerol inclusions that are initially spherical in shape. The first method yields a random distribution of inclusions, with radii ranging from $A=35$ $\mu$m to $100$ $\mu$m, that are most often far away from each other, but that can also be near one another in  clusters. On the other hand, the second method yields films where the locations of all the inclusions are known by construction. Because of the use of a needle to inject the glycerol manually into the PDMS, the radii of the injected inclusions are larger than those obtained with the first method and, in particular, \emph{not} smaller than $A=250$ $\mu$m.

\subsection{Specimens}

Given any fabricated film, owing to the transparency of PDMS and the fluorescent dye in the glycerol, the initial shape, size, and spatial location of the underlying inclusions can be expediently probed via confocal laser scanning fluorescent microscopy. This allows to identify regions in the film that contain isolated inclusions or pairs of inclusions that are far away from other inclusions. Once such regions of the film have been identified, they can be cut out into proper specimens for testing to carry out the mechanical experiments described in the next section. 

\begin{figure}[t!]
\centering
\centering\includegraphics[width=0.9\linewidth]{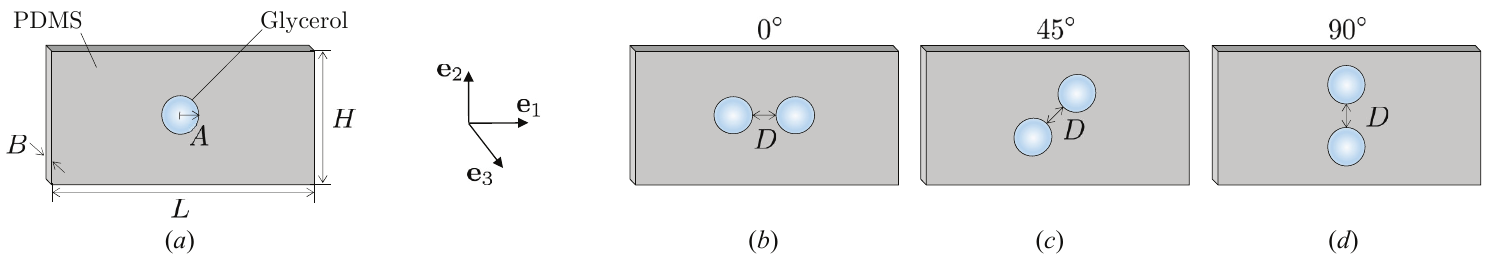}
\caption{\small Schematics of the four types of specimens tested: ($a$) specimens containing an isolated inclusion of initial radius $A$ and ($b$)--($d$) specimens containing pairs of inclusions of initial radius $A$ that are separated by a small initial distance $D<2A$ and oriented at $0^\circ$, $45^\circ$, and $90^\circ$ with respect to the $\bfe_1$ direction (the loading direction). All four types of specimens are of the same length $L=30$ mm, width $H=4$ mm, and thickness $B=1$ mm. The initial radius of the inclusions are in the range $A\in[35, 264]$ $\mu$m.}\label{Fig2}
\end{figure}

As schematically depicted in Fig. \ref{Fig2}, all the regions of interest that were cut out from the fabricated films --- with the help of single edge razor blade --- for this study were in the form of rectangular specimens of length $L=30$ mm and width $H=4$ mm in the $\bfe_1$ and $\bfe_2$ directions, respectively, and thickness $B=1$ mm in the $\bfe_3$ direction. Here, $\{\bfe_1, \bfe_2, \bfe_3\}$ stands for the laboratory frame of reference. We place its origin at the center of the specimens. 

In particular, as illustrated in Fig. \ref{Fig2}, we considered four types of specimens: ($a$) specimens with isolated inclusions and ($b$)--($d$) specimens with pairs of inclusions, separated by a small initial distance $D<2A$, that feature three different orientations of the two inclusions with respect to the $\bfe_1$ direction, $0^\circ$, $45^\circ$, and $90^\circ$. 

\subsection{Experiments}

The objective of this part of the investigation was to subject the specimens described above to a macroscopic uniaxial tensile load and to measure the resulting large deformation of the underlying liquid inclusions. To this end, an apparatus was built with the capability to stretch the specimens up to large macroscopically uniform stretches, in excess of $\ol=3$. Load cells with a 100-g force capacity (FUTEK FSH03870) were used to measure the forces applied to the specimens. The apparatus was designed to be accommodated under a fluorescence confocal microscope (Zeiss LSM 710), so that  microscale 3D imaging of the fluorescent glycerol inclusions could be carried out during the tests. The apparatus was also designed to fit under an optical microscope (Olympus Research) so as to be able to obtain \emph{in situ} full-field imaging of the specimen gauge section. 

\begin{figure}[b!]
\centering
\centering\includegraphics[width=0.99\linewidth]{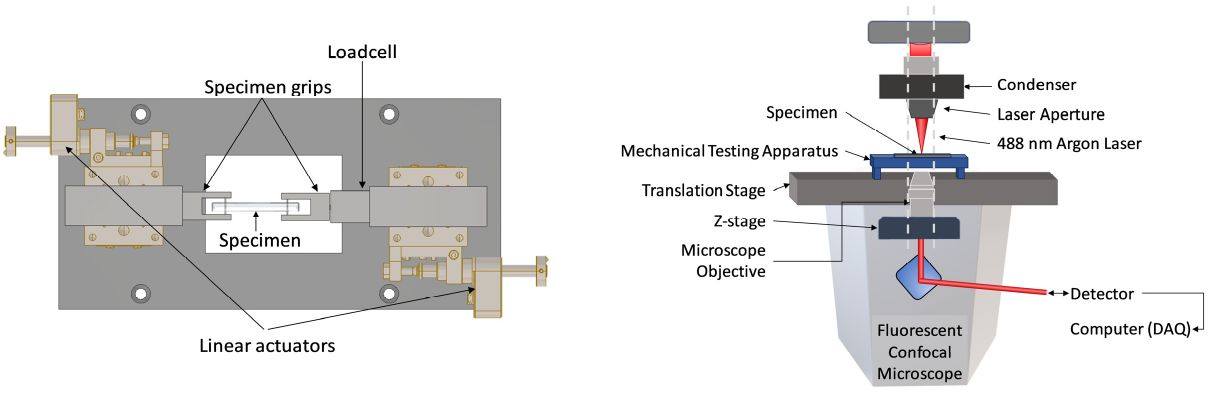}
\caption{\small Schematic of the apparatus built for the \emph{in situ} uniaxial tension experiments and test setup with the apparatus inserted under the Zeiss LSM 710 confocal microscope.}\label{Fig3}
\end{figure}
\begin{figure}[t!]
\centering
\centering\includegraphics[width=0.75\linewidth]{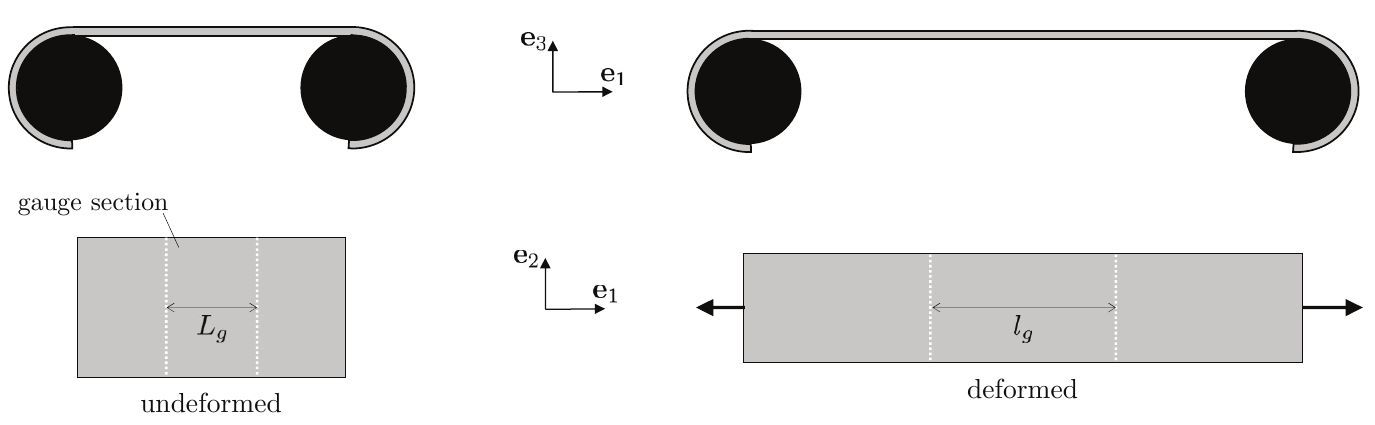}
\caption{\small Schematics of the gauge section, in the underformed and the deformed configurations, where digital image correlation (DIC) is used to measure the macroscopic stretch $\ol$ applied to the specimens. The results indicate that the stretch is macroscopically uniform --- and hence simply given by $\ol=l_g/L_g$ --- over gauge sections of initial length $L_g\approx10$ mm. }\label{Fig4}
\end{figure}

Figures \ref{Fig3} and \ref{Fig4} provide schematics of the apparatus, the setup of the tests, and the gauge section.  To carry out the experiments, the ends of the specimens were rolled around the $\bfe_2$ direction onto two Cu cylinders which, in turn, were inserted into the apparatus grips. The apparatus incorporated two linear piezoelectric actuators (New Focus 8301NF) that provided symmetric extension of the specimens such that their center remained in the optical field of view throughout the tests. All fluorescent images of liquid inclusions were taken from inclusions at the center of specimens to ensure uniform deformation and allow for easy identification of the inclusions after each loading step due to the motion of the grips (by means of symmetric stretching). A long working distance (34 mm), infinity-corrected, apochromatic objective (Mitutoyo, M Plan Apo L, $10\times$ magnification) was used for fluorescent imaging via the confocal microscope. A precise translation stage integrated with the confocal microscope was used for precise repositioning of the liquid inclusion to the center of the field of view after each loading step. All high resolution ($2048\times2048$ pixels) fluorescence images were obtained at the midplane of the glycerol inclusions where the major and minor axes were the largest and the ``halo'' at the inclusion boundary was minimized. 

Prior to their testing, all specimens were subjected to an initial preconditioning loading-unloading cycle reaching the macroscopic stretch of $\ol=1.6$. After their preconditioning, the specimens were subjected to macroscopic stretch increments of roughly $\Delta\ol= 0.03$, which were applied at a constant rate of about $\dot{\ol}=10^{-2}$ s$^{-1}$. Between increments, the stretch was held fixed for 2 min so as to carry out the confocal imaging of the underlying inclusions. Most specimens were loaded in this manner until failure, while a few were subjected to cyclic loading and unloading. Digital image correlation (Vic-2D, Correlated Solutions) was used to measure the displacement field on the surface of the gauge section of the specimens. The results indicated that the stretch was macroscopically uniform over gauge sections of initial length $L_g\approx10$ mm; see Fig. \ref{Fig4}.

\section{Theoretical description of the experiments}\label{Sec:Theory}

In order to analyze the experiments outlined above, we carried out their full-field simulation by making use of the framework introduced by \cite{GLP22}. This framework allows to describe the mechanical behaviors of the PDMS elastomer as nonlinear elastic with any hyperelastic model of choice, the glycerol inclusions as any nonlinear elastic fluid of choice, with any residual hydrostatic stress of choice, and  the PDMS/glycerol interfaces with any surface tension of choice, constant, linear, or nonlinear. 

In this work, we describe the mechanical behavior of the PDMS elastomer as hyperelastic with the non-Gaussian isotropic incompressible stored-energy function \citep{LP10}
\begin{equation}\label{LP-model}
W_{\texttt{m}}(\bfF)=\left\{\begin{array}{ll}\Psi_{\texttt{m}}(I_1) & {\rm if}\quad J=1\vspace{0.2cm}\\
+\infty & {\rm else}\end{array}\right.\quad {\rm with}\quad \Psi_{\texttt{m}}(I_1)=\displaystyle\sum_{r=1}^2\dfrac{3^{1-\alpha_r}}{2\alpha_r}\mu_r\left(I^{\alpha_r}_1-3^{\alpha_r}\right).
\end{equation}
Here, $\bfF$ denotes the deformation gradient tensor, $I_1=\bfF\cdot\bfF=\lambda_1^2+\lambda_2^2+\lambda_3^2$, $J=\det\bfF=\lambda_1\lambda_2\lambda_3$, where $\lambda_1$, $\lambda_2$, $\lambda_3$ stand for the principal stretches, while $\mu_1$, $\mu_2$, $\alpha_1$, $\alpha_2$ are material constants. The stored-energy function (\ref{LP-model}) has been previously shown to accurately describe and predict the behavior of a variety of elastomers, including PDMS Sylgard 184 with several cross-link densities \citep{PLLPR17,LWLPN20}, thus its use here.
 
Furthermore, we describe the mechanical behavior of the liquid glycerol inclusions as that of an elastic fluid with incompressible stored-energy function \citep{Truesdell73,GLP22}

\begin{equation}\label{Elastic-fluid}
W_{\texttt{i}}(\bfF)=\left\{\begin{array}{ll}\Psi_{\texttt{i}}(J)=r_{\texttt{i}} & {\rm if}\quad J=1\vspace{0.2cm}\\
+\infty & {\rm else}\end{array}\right.,
\end{equation}
where $r_{\texttt{i}}$ stands for the residual hydrostatic stress  that the liquid glycerol may be subjected to in its initial configuration (i.e., its initial pressure). 

Finally, we describe the mechanical behavior of the elastomer/glycerol interfaces as hyperelastic with constant surface tension and hence interface stored-energy function \citep{Steinmann13,GLP22}
\begin{equation}\label{Interface}
\widehat{W}(\widehat{\bfF})=\gamma\,\widehat{J}.
\end{equation}
Here, $\widehat{\bfF}=\bfF\widehat{\bfI}$ denotes the interface deformation gradient, where $\widehat{\bfI}=\bfI-\widehat{\bfN}\otimes\widehat{\bfN}$ stands for the projection tensor associated with the initial interface unit normal $\widehat{\bfN}$, pointing outwards from the glycerol inclusion towards the elastomer, $\widehat{J}=\widehat{\det}\, \widehat{\bfF}=|J\bfF^{-T}\widehat{\bfN}|$, while $\gamma$ is a material constant, to wit, the surface tension.

Given the constitutive descriptions (\ref{LP-model})--(\ref{Interface}) for the PDMS elastomer, the liquid glycerol inclusions, and their interface, neglecting inertia and body forces, the Lagrangian equations of equilibrium for the experiments under study here specialize to the system of coupled partial differential equations \citep{GLP22}
\begin{equation}\label{Equilibrium equations}
\left\{\begin{array}{ll}{\rm Div}\,\bfS=\mathbf{0}, & \bfX\in\mathrm{\Omega}_0\setminus \mathrm{\Gamma}_0\vspace{0.2cm}\\
\det\nabla\bfy=1, & \bfX\in\mathrm{\Omega}_0\setminus  \mathrm{\Gamma}_0\vspace{0.2cm}\\
\widehat{{\rm Div}}\,\widehat{\bfS}-\left\llbracket \bfS \right\rrbracket\widehat{\bfN}=\mathbf{0}, & \bfX\in \mathrm{\Gamma}_0\vspace{0.2cm}\\
\bfy=\overline{\bfy}, & \bfX\in\partial\mathrm{\Omega}^{\mathcal{D}}_0\vspace{0.2cm}\\
\bfS\bfN=\overline{\bfs}, & \bfX\in\partial\mathrm{\Omega}^{\mathcal{N}}_0\vspace{0.2cm}\\ \end{array}\right.
\end{equation}
for the deformation field $\bfy(\bfX)$ and the pressure field $p(\bfX)$, where
\begin{align}\label{Bulk-stress}
\bfS(\bfX)&=2(1-\theta_{\texttt{i}}(\bfX))\dfrac{{\rm d}\Psi_{\texttt{m}}}{{\rm d} I_1}(I_1)\nabla\bfy-p\nabla\bfy^{-T}\nonumber\\
&=(1-\theta_{\texttt{i}}(\bfX))\left(\displaystyle\sum_{r=1}^2 3^{1-\alpha_r}\mu_r I_1^{\alpha_r-1}\right)\nabla\bfy-p\nabla\bfy^{-T},\quad \bfX\in \mathrm{\Omega}_0\setminus  \mathrm{\Gamma}_0
\end{align}
and
\begin{equation}\label{Interface-stress}
\widehat{\bfS}(\bfX)=\gamma\,\widehat{\det}\,(\widehat{\nabla}{\bfy})\widehat{\nabla}{\bfy}^{-T},\quad \bfX\in\mathrm{\Gamma}_0
\end{equation}
are the first Piola-Kirchhoff stress tensors in the bulk and at the interfaces. In these expressions, 
\begin{equation}
\theta_{\texttt{i}}(\bfX)=\left\{\begin{array}{ll}
1 & \text{if}\quad \bfX\in \mathrm{\Omega}_0^{(\texttt{i})}\\
0 & \text{else} \end{array}\right. \label{thetai}
\end{equation}
denotes the characteristic or indicator function that identifies the domain $\mathrm{\Omega}_0^{(\texttt{i})}\subset \mathrm{\Omega}_0$ occupied by the inclusions within the domain $\mathrm{\Omega}_0$ occupied by the specimens in their initial configuration, $\mathrm{\Gamma}_0$ denotes the interfaces that separate the elastomer from the inclusions, while $\partial\mathrm{\Omega}^{\mathcal{D}}_0$ and $\partial\mathrm{\Omega}^{\mathcal{N}}_0=\partial\mathrm{\Omega}_0\setminus\partial\mathrm{\Omega}^{\mathcal{D}}_0$ denote the Dirichlet and Neumann parts of the boundary $\partial\mathrm{\Omega}_0$ where the deformation $\overline{\bfy}$ and traction $\overline{\bfs}$ are prescribed. Moreover, $\llbracket \cdot \rrbracket$  is the jump operator across the interfaces $\mathrm{\Gamma}_0$ based on the convention $\llbracket f(\bfX) \rrbracket=f_{\texttt{i}}-f_{\texttt{m}}$, where $f_{\texttt{i}}$ (resp. $f_{\texttt{m}}$) denotes the limit of any given function $f(\bfX)$ when approaching $\mathrm{\Gamma}_0$ from within the inclusion (resp. elastomer), while $\widehat{\nabla}$ and $\widehat{{\rm Div}}$ stand for the interface gradient and divergence operators. That is, in terms of the projection tensor $\widehat{\bfI}$ and indicial notation, $(\widehat{\nabla}\bfv)_{ij}=(\partial v_i/\partial X_k) \widehat{I}_{kj}$ and $(\widehat{{\rm Div}}\,\bfT)_{i}=(\partial T_{ij}/\partial X_k) \widehat{I}_{kj}$ when applied to a vector $\bfv$ and a second-order tensor $\bfT$.

\begin{figure}[b!]
\centering
\centering\includegraphics[width=0.99\linewidth]{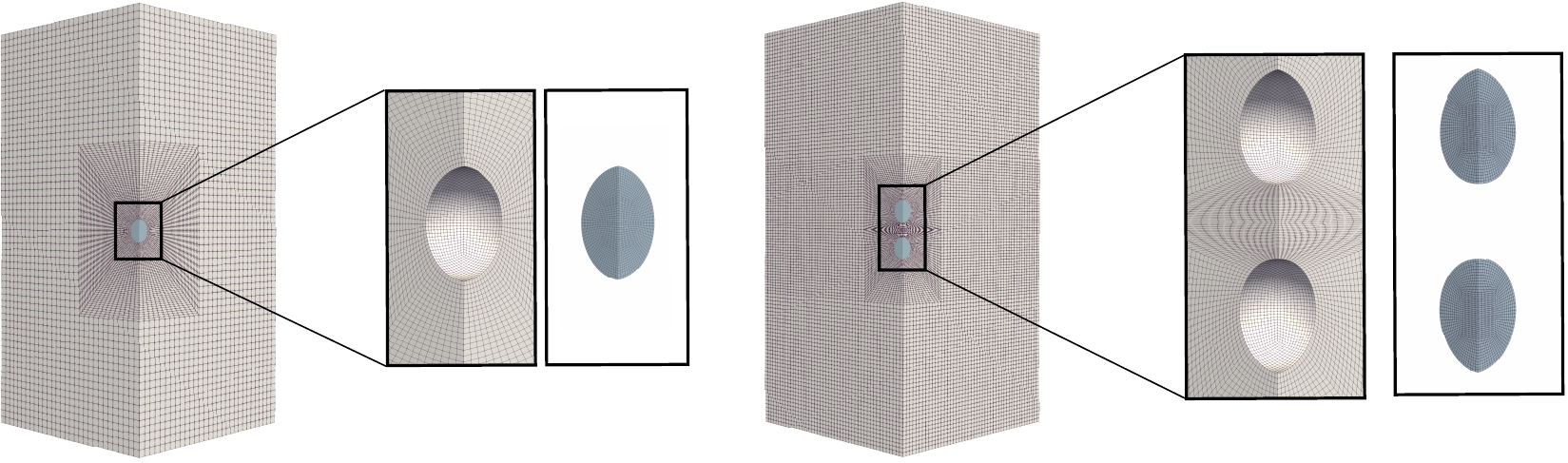}
\caption{\small Examples of the FE discretizations used for the problem with an isolated liquid inclusion and for that of a pair of inclusions oriented at $0^\circ$ with respect to the $\bfe_1$ direction (the loading direction).}\label{Fig5}
\end{figure}

Now, numerical experiments show that the boundaries of the specimens can be considered to be infinitely far away from the inclusions. In view of this and the fact that the macroscopic deformation in the gauge section is uniform, we take the domain $\mathrm{\Omega}_0$ occupied by any specimen in its initial configuration to be a cube of initial side $40A$, specifically,
\begin{equation}\label{Omega}
\mathrm{\Omega}_0=\left\{\bfX: -20 A<X_i<20A\quad i=1,2,3\right\}.
\end{equation}
A domain of this kind is large enough to yield solutions that effectively can be considered as solutions for inclusions embedded in an infinite domain; cf. \cite{LPGD13a} and \cite{LDLP17}. Furthermore, we take the domain $\mathrm{\Omega}_0^{(\texttt{i})}$ occupied by the isolated liquid inclusions of radius $A$ in their initial configuration to be given by
\begin{equation}\label{Omegai}
\mathrm{\Omega}_0^{(\texttt{i})}=\left\{\bfX: |\bfX|<A\right\}
\end{equation}
and that occupied by the pairs of liquid inclusions of radius $A$, separated by a distance $D$, to be given by
\begin{equation}\label{Omegai-pair}
\mathrm{\Omega}_0^{(\texttt{i})}=\left\{\bfX: |\bfX-\bfX^{(1)}|<A,\,|\bfX-\bfX^{(2)}|<A\right\}\quad {\rm with}\quad \left\{\begin{array}{l} 0^\circ\left\{\begin{array}{l}X_1^{(1)}=-\frac{D}{2}-A,\,X_2^{(1)}=0,\, X_3^{(1)}=0\vspace{0.2cm}\\X_1^{(2)}=\frac{D}{2}+A,\,X_2^{(2)}=0,\, X_3^{(2)}=0\end{array}\right.\vspace{0.2cm}\\ 45^\circ\left\{\begin{array}{l}X_1^{(1)},X_2^{(1)}=-\frac{1}{\sqrt{2}}\left(\frac{D}{2}+A\right),\, X_3^{(1)}=0\vspace{0.2cm}\\X_1^{(2)},X_2^{(2)}=\frac{1}{\sqrt{2}}\left(\frac{D}{2}+A\right),\, X_3^{(2)}=0\end{array}\right. \vspace{0.2cm}\\ 90^\circ\left\{\begin{array}{l}X_1^{(1)}=0,\,X_2^{(1)}=-\frac{D}{2}-A,\, X_3^{(1)}=0\vspace{0.2cm}\\X_1^{(2)}=0,\,X_2^{(2)}=\frac{D}{2}+A,\, X_3^{(2)}=0\end{array}\right. \vspace{0.2cm}\\ \end{array}\right. ;
\end{equation}
see Fig. \ref{Fig2}. By the same token, it suffices to consider boundary conditions of the affine form
\begin{equation}
\left\{\hspace{-0.05cm}\begin{array}{ll}
\overline{y}_1(\bfX)=20A \ol, & \bfX\in\partial\mathrm{\Omega}^{\mathcal{T}}_0\vspace{0.2cm}\\
\overline{s}_2(\bfX)=\overline{s}_3(\bfX)=0, & \bfX\in\partial\mathrm{\Omega}^{\mathcal{T}}_0\vspace{0.2cm}\\
\overline{y}_1(\bfX)=-20A \ol, & \bfX\in\partial\mathrm{\Omega}^{\mathcal{B}}_0\vspace{0.2cm}\\
\overline{s}_2(\bfX)=\overline{s}_3(\bfX)=0, & \bfX\in\partial\mathrm{\Omega}^{\mathcal{B}}_0\vspace{0.2cm}\\
\overline{\bfs}(\bfX)=\textbf{0}, & \bfX\in\partial\mathrm{\Omega}_0\setminus(\partial\mathrm{\Omega}^{\mathcal{T}}_0\cup\partial\mathrm{\Omega}^{\mathcal{B}}_0)
\end{array}\right.,\label{BCs}
\end{equation}
where $\partial\mathrm{\Omega}^{\mathcal{T}}_0=\{\bfX: X_1=20A,\;-20A<X_2,X_3<20A\}$ and $\partial\mathrm{\Omega}^{\mathcal{B}}_0=\{\bfX: X_1=-20A,\;-20A<X_2,X_3<20A\}$ and, again, $\ol=l_g/L_g$ stands for the macroscopic stretch applied in the experiments; see Fig. \ref{Fig4}.

Given the domains (\ref{Omega})-(\ref{Omegai-pair}) and boundary conditions (\ref{BCs}), the resulting PDEs (\ref{Equilibrium equations}) can only be solved numerically. We do so by means of a modified version of the FE (finite element) scheme introduced by \cite{GLP22}. Figure \ref{Fig5} presents two examples of the structured FE dicretizations used to generate such solutions.

\section{Results and discussion}\label{Sec:Results}

\subsection{The nonlinear elastic response of the PDMS elastomer}

We begin by presenting the results that describe the mechanical behavior of the PDMS elastomer used to fabricate the specimens. Several specimens of pure PDMS, without inclusions, were fabricated and tested under uniaxial tension at a constant stretch rate of about $|\dot{\ol}|=10^{-2}$ s$^{-1}$ in the apparatus shown in Fig. \ref{Fig3}. Some of these tests were carried out until failure of the specimen, while others were carried out under cyclic loading and unloading.

\begin{figure}[H]
\centering
\centering\includegraphics[width=0.375\linewidth]{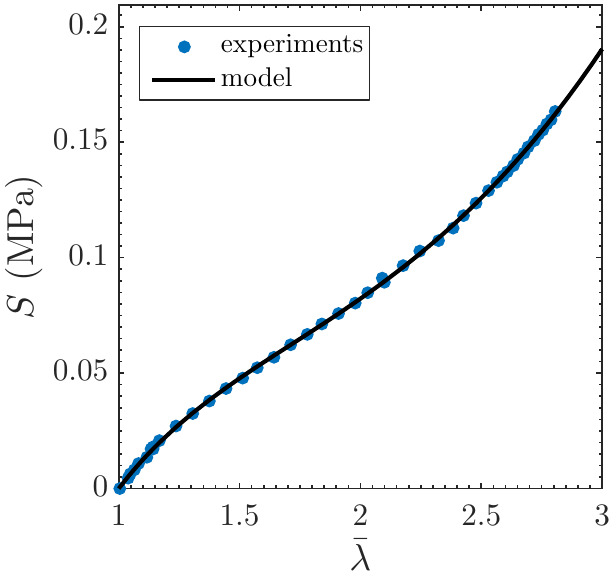}
\caption{\small Uniaxial tension response of the PDMS elastomer used in the experiments. The results show the nominal stress $S$ versus the applied stretch $\ol$ from three different experiments, as well as the theoretical stress-stretch response (\ref{S-LP}) described by the model (\ref{LP-model}) with the material constants listed in Table \ref{Table1}.}\label{Fig6}
\end{figure}

\begin{table}[H]\centering
\caption{Fitted values of the material constants in the hyperelastic model (\ref{LP-model}) for the PDMS elastomer used in the experiments.}
\begin{tabular}{cc|cc}
\toprule
$\mu_1$ (kPa)  & $\alpha_1$ & $\mu_2$ (kPa) & $\alpha_2$\\
\hline
$33.5$   & $0.3648$ & $12$  & $2.1897$\\
\bottomrule
\end{tabular} \label{Table1}
\end{table}
Consistent with data reported in the literature for PDMS Sylgard 184 with a similar weight ratio (30:1) of elastomer to curing agent \citep{PLLPR17,LWLPN20}, the results show that the mechanical response of the elastomer is essentially that of an isotropic nearly incompressible nonlinear elastic material. Figure \ref{Fig6} presents the average stress-stretch response obtained from three different tests. The figure also includes the stress-stretch response 
\begin{equation}\label{S-LP}
S=\left(\ol-\ol^{\,-2}\right)\displaystyle\sum_{r=1}^2 3^{1-\alpha_r}\mu_r\left(\ol^2+2\ol^{\,-1}\right)^{\alpha_r-1}
\end{equation}
described by the model (\ref{LP-model}) with the material constants listed in Table \ref{Table1}, which were obtained by fitting (\ref{S-LP}) to the experimental data. For later reference, we note here that these material constants imply that the initial shear modulus of the elastomer is $\mu=\mu_1+\mu_2=45.5$ kPa. It is plain from Fig. \ref{Fig6} that the model (\ref{LP-model}) with such material constants describes reasonably well the experimentally measured response. 

\subsection{The surface tension $\gamma$ at the elastomer/inclusion interfaces and the residual stress $r_{\texttt{i}}$ in the inclusions}

In addition to the material constants listed in Table \ref{Table1} that describe the elastic behavior of the elastomer, there are two other material properties that enter the governing equations (\ref{Equilibrium equations}): the surface tension $\gamma$ at the elastomer/inclusion interfaces and the residual stress $r_{\texttt{i}}$ in the inclusions. We did not measure these properties directly, instead, we estimated their values from earlier works in the literature.

Specifically, from experiments on glycerol inclusions embedded in a similar type of silicone elastomer \citep{Syleetal15}, the surface tension at the elastomer/inclusion interfaces in our specimens is estimated to be about
\begin{equation}\label{gamma-exp}
\gamma=0.014\,{\rm N/m},
\end{equation}
at least initially, when the inclusions are not exceedingly deformed. Making use of this estimate, it follows from the PDEs (\ref{Equilibrium equations})$_{1,3}$ that the residual hydrostatic stress in the inclusions should be in the range
\begin{equation}\label{ri-exp}
r_{\texttt{i}}=-\dfrac{2\gamma}{A}\in[-0.8, -0.1] \, {\rm kPa}\quad {\rm for }\quad A\in[35, 264]\, \mu{\rm m}.
\end{equation}
By the same token, given that the initial shear modulus of the PDMS elastomer is $\mu=45.5$ kPa, the initial elasto-capillary number is estimated to be in the range
\begin{equation}\label{eCa-exp}
eCa=\dfrac{\gamma}{2\mu A}\in[5.8\times 10^{-4}, 4.4\times 10^{-3}] \quad {\rm for }\quad A\in[35, 264]\, \mu{\rm m}.
\end{equation}
The values (\ref{gamma-exp})--(\ref{eCa-exp}) are small enough that the presence of surface tension at the elastomer/inclusion interfaces and residual hydrostatic stress in the inclusions should be negligible \citep{GLLP23b}. This is indeed confirmed by direct simulations, which show that results for $\gamma=0$ and $r_{\texttt{i}}=0$ are essentially identical to those for $\gamma=0.014$ N/m and $r_{\texttt{i}}\in[-0.8,-0.1]$ kPa.

\subsection{The deformation of isolated liquid inclusions}

\begin{figure}[t!]
\centering
\centering\includegraphics[width=0.9\linewidth]{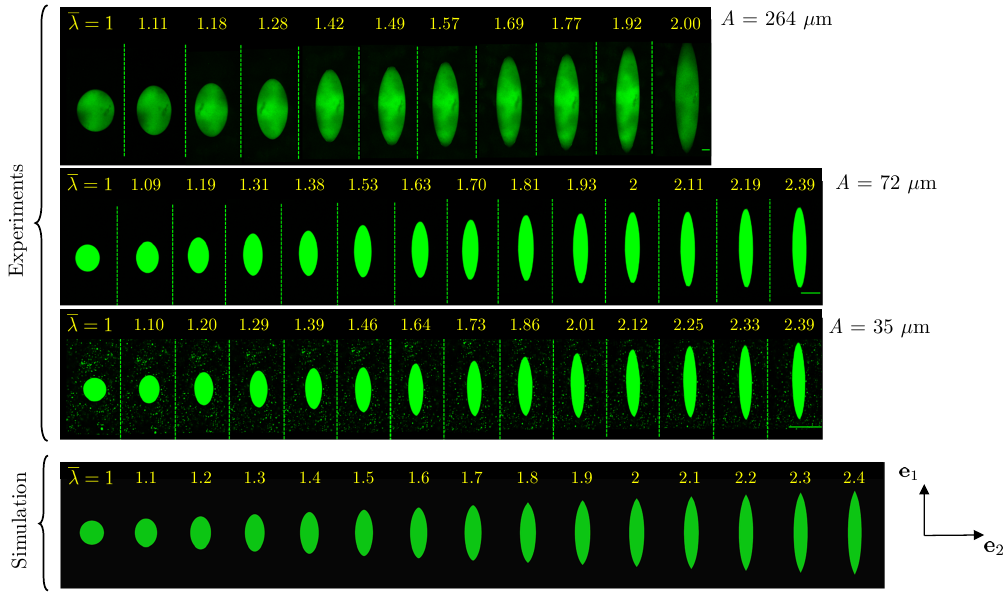}
\caption{\small  Fluorescent confocal microscopy images showing the deformation across the midplane of isolated inclusions at increasing values of applied macroscopic stretch $\ol$. The experimental results pertain to three different specimens with inclusions of three different initial radii, $A=35, 72, 264$ $\mu$m; scale bars are 100 $\mu$m. For direct comparison, the figure includes the corresponding results from a simulation.}\label{Fig7}
\end{figure}
\begin{figure}[t!]
\centering
\centering\includegraphics[width=0.85\linewidth]{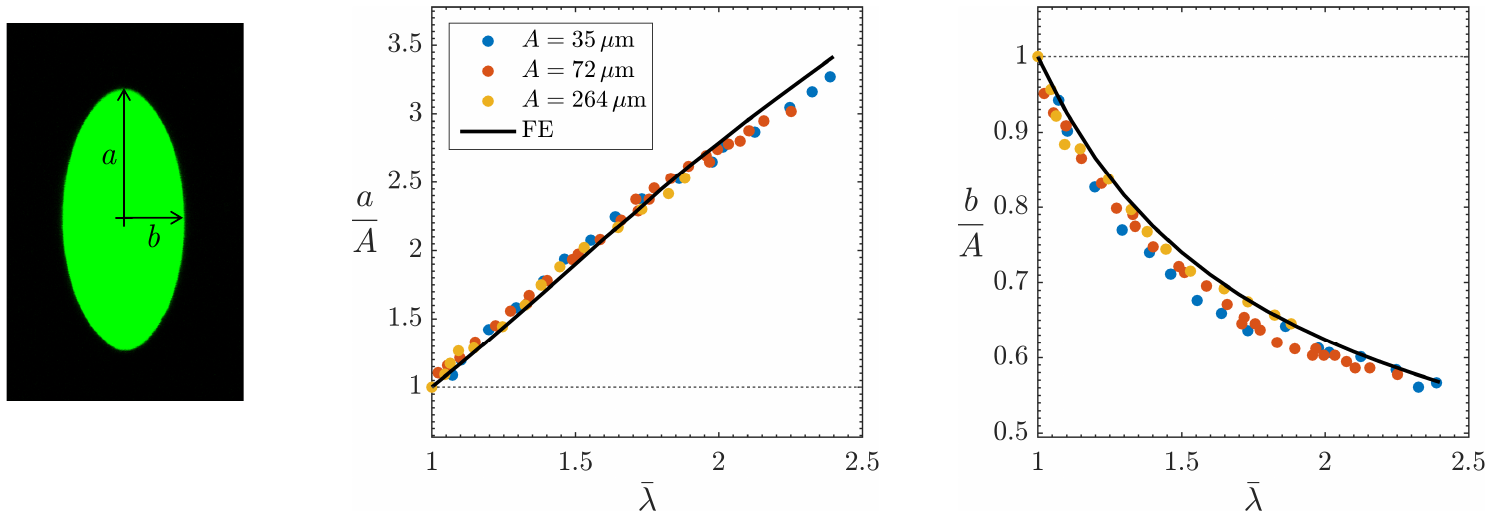}
\caption{\small Evolution of the major, $a$, and minor, $b$, semi-axes of the isolated inclusions shown in Fig. \ref{Fig7}. The results are shown normalized by the initial radii $A$ of the inclusions, as a function of the applied macroscopic stretch $\ol$.}\label{Fig8}
\end{figure}
\begin{figure}[t]
\centering
\centering\includegraphics[width=0.85\linewidth]{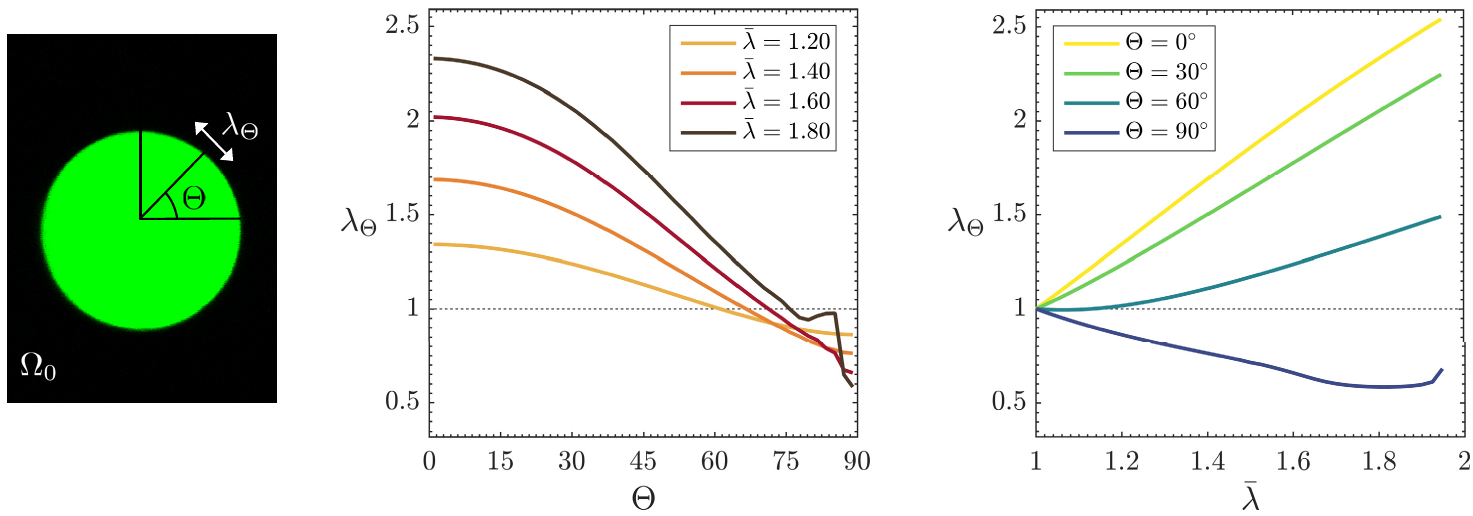}
\caption{\small The hoop stretch $\lambda_{\Theta}$ along the elastomer/inclusion interface in the simulation for the isolated inclusion shown in Fig. \ref{Fig7}. The results are shown for several values of the applied macroscopic stretch $\ol$, as a function of the hoop angle $\Theta$, and for several values of $\Theta$, as a function of $\ol$.}\label{Fig9}
\end{figure}

At this stage, we are ready to start presenting and analyzing the deformation of the inclusions. Figure \ref{Fig7} shows sequences of high-resolution fluorescent confocal microscopy images of isolated inclusions of three different initial radii, $A=35, 72, 264$ $\mu$m, as the macroscopic stretch $\ol$ applied to the specimens increases. The figure also shows the corresponding sequence obtained from a simulation.  Since the FE results are inclusion-size independent, only one simulation is presented. We remind the reader that all the images correspond to the midplane of the inclusions. 

As one might expect intuitively, the inclusions are seen to deform symmetrically, extending in the direction $\bfe_1$ of the applied macroscopic stretch $\ol$ and contracting in the perpendicular $\bfe_2$-$\bfe_3$ plane. As we anticipated from the very small values (\ref{eCa-exp}) of the elasto-capillary number $eCa$, the deformation of the inclusions appears to be independent of their size. This size-independence is confirmed by Fig. \ref{Fig8}, where the normalized sizes $a/A$ and $b/A$ of the major and minor semi-axes of the three isolated inclusions are plotted as functions of $\ol$. Figure \ref{Fig8} includes as well the corresponding results from the simulation. It is plain from Figs. \ref{Fig7} and \ref{Fig8} that the simulation is in good qualitative and quantitative agreement with the experiments. 

Beyond illustrating the size-independent character of the deformation of the inclusions in the experiments and the agreement between those and the simulation, the results in Fig. \ref{Fig8} serve to illustrate how much more the inclusions stretch relative to the applied macroscopic stretch $\ol$. Indeed, it is clear from the plots that $a/A>\ol$ and $b/A<\ol^{\,-1/2}$. This is a direct consequence of the fact that the inclusions are ``softer'' than the surrounding elastomer because they are made of a liquid, which exhibits no shear resistance, and because the ``stiffening'' provided by the surface tension at the elastomer/inclusion interfaces is negligible. 

To gain further insight into the deformation of the inclusions beyond the measures $a/A$ and $b/A$, it proves useful to examine the hoop stretch
\begin{equation}\label{l-Theta}
\lambda_{\Theta}:=\left.\sqrt{\bfF(\bfX)\bfv_{\Theta}\cdot\bfF(\bfX)\bfv_{\Theta}}\right\vert_{|\bfX|=A} \quad {\rm with}\quad\bfv_{\Theta}=\cos\Theta\,\bfe_1-\sin\Theta\,\bfe_2
\end{equation}
along the elastomer/inclusion interface. Figure \ref{Fig9} presents results for $\lambda_{\Theta}$ obtained from the simulation in Fig. \ref{Fig7}  for several values of the applied macroscopic stretch $\ol$, as a function of the hoop angle $\Theta$, and for several values of $\Theta$, as a function of $\ol$. These results call for the following observations. Save for a localized $\ol$-dependent region near the north pole, at $\Theta=\pi/2$, the hoop stretch is tensile ($\lambda_{\Theta}>1$) and decreases monotonically with increasing $\Theta$, while it increases monotonically with increasing $\ol$. Accordingly, at any given macroscopic stretch $\ol$, the largest hoop stretch is always attained at the equator, when $\Theta=0$. As $\Theta$ increases and the pole $\Theta=\pi/2$ is approached, the hoop stretch eventually changes to be compressive ($\lambda_{\Theta}<1$). As elaborated in the next subsection, sufficiently large compressive hoop stretches lead to the development of a crease at the poles of the inclusion.

\begin{figure}[H]
\centering
\centering\includegraphics[width=0.85\linewidth]{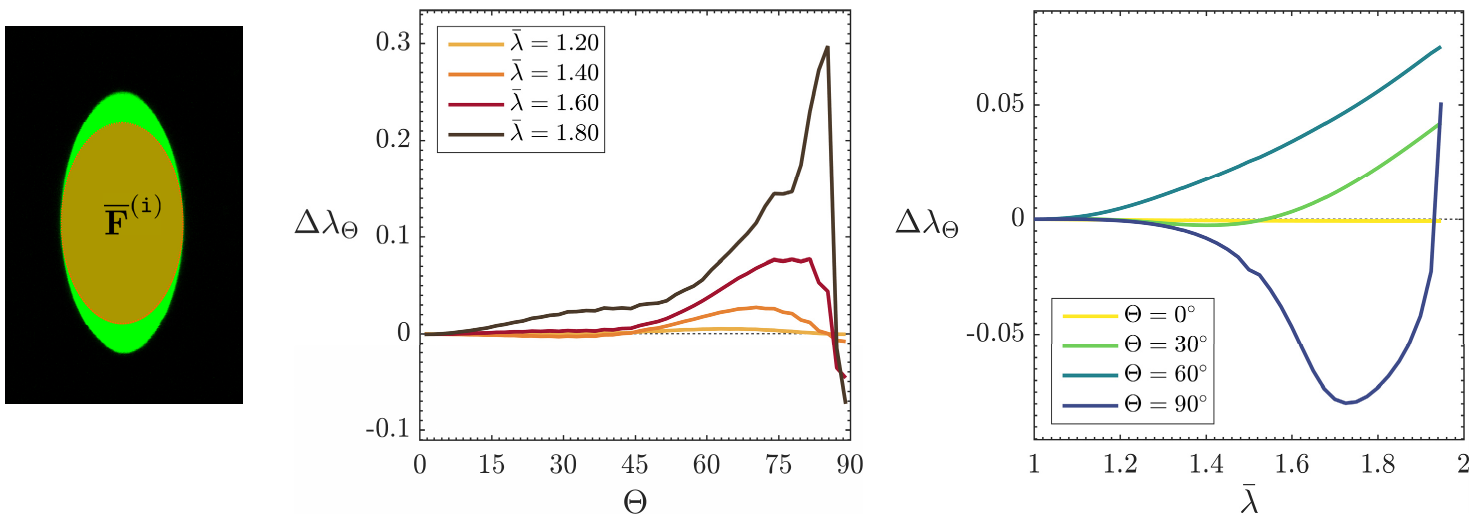}
\caption{\small Difference $\Delta\lambda_{\Theta}=\lambda_{\Theta}-\ol_{\texttt{i}_{\Theta}}$ between the hoop stretch $\lambda_{\Theta}$ along the elastomer/inclusion interface in the simulation for the isolated inclusion shown in Fig. \ref{Fig7} and the hoop stretch $\ol_{\texttt{i}_{\Theta}}$ on an inclusion of perfect ellipsoidal shape with constant deformation gradient (\ref{Fi}). The results are shown for several values of the applied macroscopic stretch $\ol$, as a function of the hoop angle $\Theta$, and for several values of $\Theta$, as a function of $\ol$.}\label{Fig10}
\end{figure}

We close this subsection by quantifying, in terms of the hoop stretch $\lambda_{\Theta}$, how different the deformation in the isolated liquid inclusions is from a uniform deformation. It is well known from the work of \cite{Eshelby57} that the deformation gradient within an isolated liquid inclusion of the type of interest here is uniform for macroscopic stretches $\ol$ sufficiently close to $\ol=1$ and that this implies that the inclusion deforms into an ellipsoidal shape. However, as $\ol$ deviates away from $\ol=1$, the deformation gradient within the inclusion will become non-uniform and, in turn, the shape of the inclusion will deviate from that of an ellipsoid. To quantify this non-uniformity and deviation from perfect ellipsoidal shape, we make use of the measure
\begin{equation}\label{Delta l-Theta}
\Delta\lambda_{\Theta}=\lambda_{\Theta}-\ol_{\texttt{i}_{\Theta}}
\end{equation}
in terms of the hoop stretch
\begin{equation*}
\ol_{\texttt{i}_{\Theta}}:=\sqrt{\obfF^{(\texttt{i})}\bfv_{\Theta}\cdot\obfF^{(\texttt{i})}\bfv_{\Theta}}=\ol_{\texttt{i}}\sqrt{\cos^2\Theta+\dfrac{\sin^2\Theta}{\ol^{\,3}_{\texttt{i}}}}
\end{equation*}
associated with the constant deformation gradient 
\begin{equation}\label{Fi}
\obfF^{(\texttt{i})}:=\ol_{\texttt{i}}\bfe_1\otimes\bfe_1+\ol^{-1/2}_{\texttt{i}}(\bfe_2\otimes\bfe_2+\bfe_3\otimes\bfe_3)\quad {\rm with}\quad \ol_{\texttt{i}}=\left.\lambda_{\Theta}\right\vert_{\Theta=0}.
\end{equation}
By definition, $\Delta\lambda_{\Theta}=0$ for all $\ol$ and all $\Theta$ whenever the deformation gradient within the inclusion is uniform. Furthermore, by definition, $\Delta\lambda_{\Theta}=0$ for all $\ol$ at $\Theta=0$. Deviations from $\Delta\lambda_{\Theta}=0$ for $\Theta\neq0$ signal then deviations from a uniform field within the inclusion.

Figure \ref{Fig10} presents results for the measure (\ref{Delta l-Theta}) based on the hoop stretch $\lambda_{\Theta}$ obtained from the simulation for the isolated inclusion shown in Fig. \ref{Fig7}. The results are shown both for several values of the applied macroscopic stretch $\ol$, as a function of the hoop angle $\Theta$, and for several values of $\Theta$, as a function of $\ol$. A quick glance suffices to recognize that the deformation of the inclusion is indeed \emph{not} uniform, more so the larger the applied macroscopic stretch $\ol$.

\subsection{Creases at the elastomer/inclusion interfaces}

\begin{figure}[b!]
\centering
\centering\includegraphics[width=0.9\linewidth]{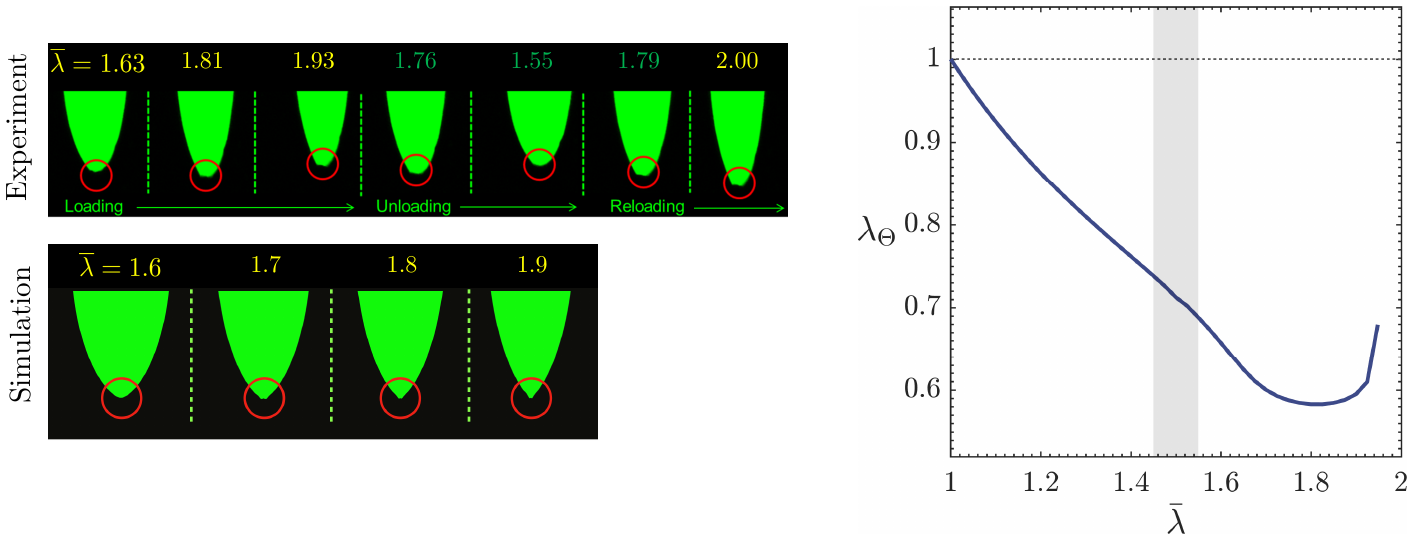}
\caption{\small  Fluorescent confocal microscopy images showing the deformation across the midplane and around the north pole ($\Theta=\pi/2$) of an isolated inclusion with initial radius $A=72$ $\mu$m at several values of the applied macroscopic stretch $\ol$. The figure includes the corresponding images from a simulation and the evolution in this simulation of the hoop stretch $\ol_{\Theta}$ at $\Theta=\pi/2$ as a function of $\ol$.}\label{Fig11}
\end{figure}

As noted in the discussion of Fig. \ref{Fig9} above, the poles of the inclusions are subjected to a compressive hoop stretch that, when large enough, can lead to the development of creases. Figure \ref{Fig11} presents a sequence of high-resolution fluorescent confocal microscopy images that zoom in the development of a crease at the north pole of an isolated inclusion of initial radius $A = 72$ $\mu$m. To examine whether the appearance of such a crease is an elastic phenomenon, images of the loading, unloading, and reloading of the specimen are included. The figure also includes the corresponding sequence obtained from a simulation, as well as a plot of the evolution in this simulation of the hoop stretch $\ol_{\Theta}$ at $\Theta=\pi/2$, the north pole, as a function of the macroscopic stretch $\ol$.

The experiment shows the appearance of a crease at a macroscopic stretch of about $\ol=1.6$. As the specimen is stretched further, the crease grows in size, while it disappears upon unloading and reappears upon reloading, indicating that the emergence of such a localized deformation is a purely elastic phenomenon. This behavior is consistent with the appearance of creases in blocks of elastomers and gels when compressed or bent. Since the classical experiments of \cite{Gent99} on blocks of rubber, numerous studies have been and continue to be devoted to understanding the nucleation of creases in soft materials; see, e.g., \cite{Suo09}, \cite{Mahadevan11}, \cite{Kim13}, \cite{Truskinovsky19}, \cite{Fu20}, and \cite{Triantafyllidis22}. Save for the related results presented in \citep{Zhao12}, however, we are not aware of any studies of creases at the interface of liquid inclusions with embedding elastomers. It is expected that the 3D dome-like geometry around the poles where the creases appear, as well as the presence of a pressurized liquid and also possibly the presence of elasto-capillary effects may have a significant impact on when creases form at such interfaces. 

For the problem at hand here, the FE results presented in Fig. \ref{Fig11} show the appearance of a crease at a macroscopic stretch of about $\ol=1.5$, which is in fair agreement with the experimentally estimated value. The corresponding critical hoop stretch at $\Theta=\pi/2$ is about $\lambda_{\Theta}=0.71$. This value is notably larger (i.e., it is a smaller compressive stretch) than the one obtained for the prototypical nucleation of a crease in a half-space made of the same PDMS elastomer under plane-strain compression in the absence of a pressurized liquid, namely, $\lambda_{c}=0.65$. Importantly, no imperfection is purposely placed in the FE mesh. Instead, it appears that the errors inherent to the use of a FE discretization are sufficient as imperfections to break the symmetry of the deformation and to lead to the formation of a crease at the pole.  

\subsection{The deformation of two interacting liquid inclusions}

\begin{figure}[b!]
\centering
\centering\includegraphics[width=0.95\linewidth]{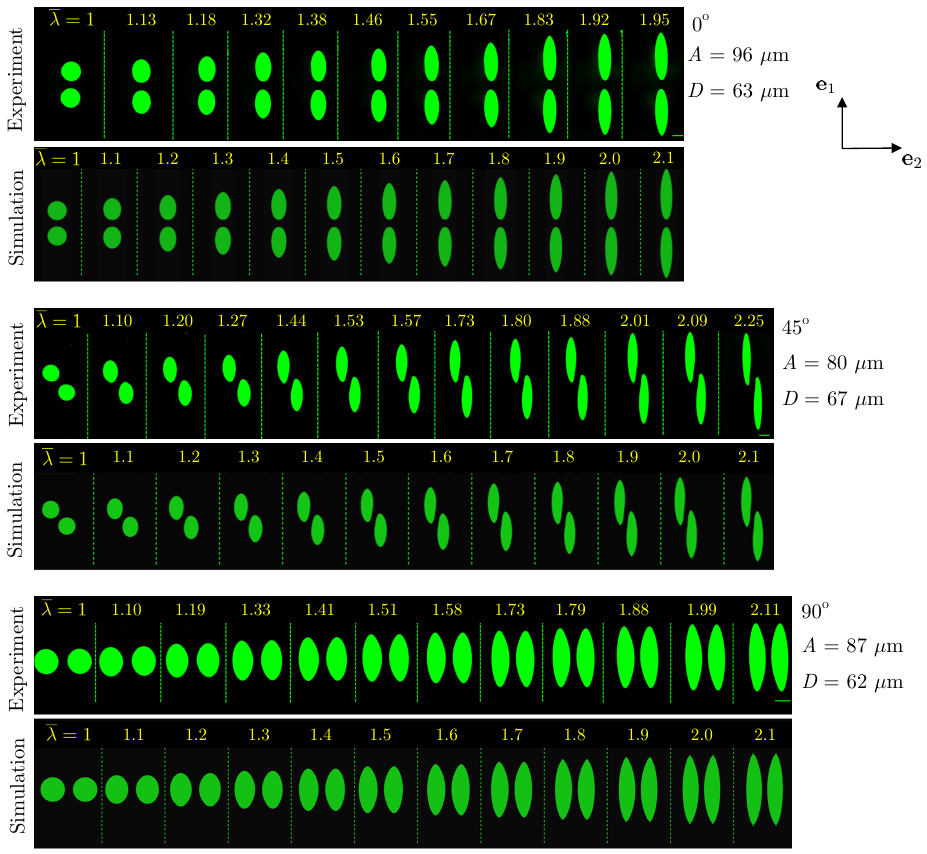}
\caption{\small  Fluorescent confocal microscopy images showing the deformation across the midplane of pairs of inclusions, oriented at $0^\circ$, $45^\circ$, and $90^\circ$ with respect to the $\bfe_1$ direction (the loading direction), at increasing values of the applied macroscopic stretch $\ol$. The results pertain, respectively, to inclusions of initial radii $A=96, 80, 87$ $\mu$m separated by initial distances $D=63,67,62$ $\mu$m; scale bars are 100 $\mu$m. For direct comparison, the figure includes the corresponding results from simulations.}\label{Fig12}
\end{figure}
\begin{figure}[t!]
\centering
\centering\includegraphics[width=0.99\linewidth]{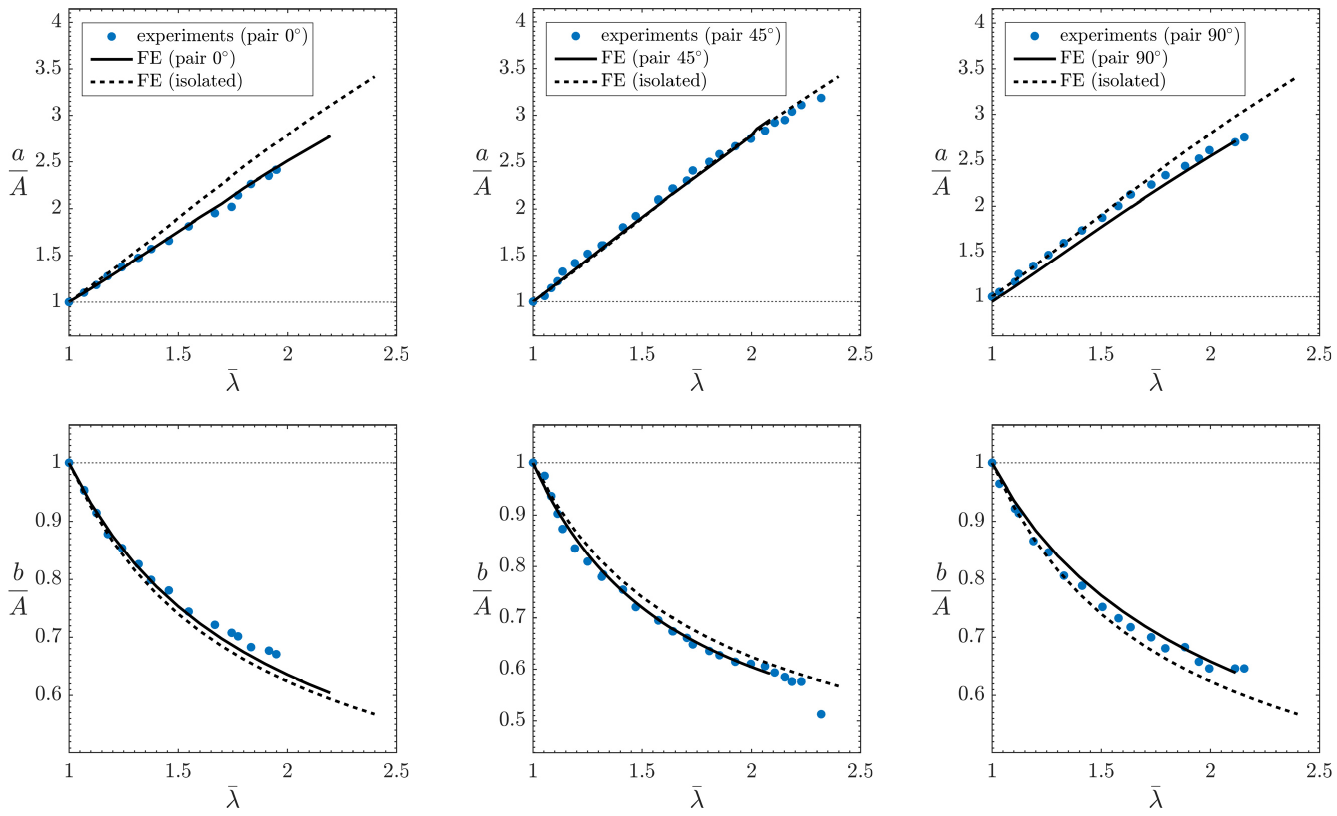}
\caption{\small Evolution of the sizes $a$ and $b$ of the major and minor semi-axes of the individual inclusions in each of the three pairs of inclusions shown in Fig. \ref{Fig12}. The results are shown normalized by the initial radii $A$ of the inclusions, as functions of the applied macroscopic stretch $\ol$. For direct comparison, the plots include the corresponding data for an isolated inclusion, as determined from simulations.}\label{Fig13}
\end{figure}

Next, we turn to the presentation and analysis of the deformation of pairs of closely interacting inclusions. Analogous to Fig. \ref{Fig7}, Fig. \ref{Fig12} shows sequences of high-resolution fluorescent confocal microscopy images of three pairs of inclusions of initial radii  $A=96, 80, 87$ $\mu$m, separated by initial distances $D=63,67,62$ $\mu$m and oriented at $0^\circ$, $45^\circ$, and $90^\circ$ with respect to the $\bfe_1$ direction (the loading direction), as the macroscopic stretch $\ol$ applied to the specimens increases. The figure also shows the corresponding sequences obtained from simulations.  

As is the case of the isolated inclusions, the pairs of inclusions are seen to extend in the direction $\bfe_1$ of the applied macroscopic stretch $\ol$ and contract in the perpendicular $\bfe_2$-$\bfe_3$ plane. The specifics of their deformation, however, are more intricate than those found for isolated inclusions and depend strongly on the orientation of the pair with respect to the loading direction. So as to quantify this difference, Fig. \ref{Fig13} presents plots, as functions of $\ol$, of the normalized sizes $a/A$ and $b/A$ of the major and minor semi-axes of the individual inclusions in each of the experiments and simulations shown in Fig. \ref{Fig12}. The figure includes as well the corresponding results for an isolated inclusion. 

All the results in Fig. \ref{Fig13} are such that $a/A>\ol$ indicating that, notwithstanding their close interaction, the inclusions behave ``softer'' than the surrounding elastomer. The inclusions that exhibit the ``softest'' response --- in the sense of exhibiting the largest $a/A$ --- are those that are oriented at $45^\circ$. The ones that are oriented at $0^\circ$, on the other hand, exhibit the ``stiffest'' response. This is due to the fact that their closest poles exhibit mutual shielding; see Fig. \ref{Fig12}. An analogous shielding effect is observed for the pair oriented at $90^\circ$, which results in a ratio $b/A$ that is larger than that of an isolated inclusion. Interestingly, even though their local deformation is significantly different from that of isolated inclusions, the inclusions oriented at $45^\circ$ exhibit an evolution of their major semi-axis that is very similar to that of isolated inclusions.

\begin{figure}[t!]
\centering
\centering\includegraphics[width=0.85\linewidth]{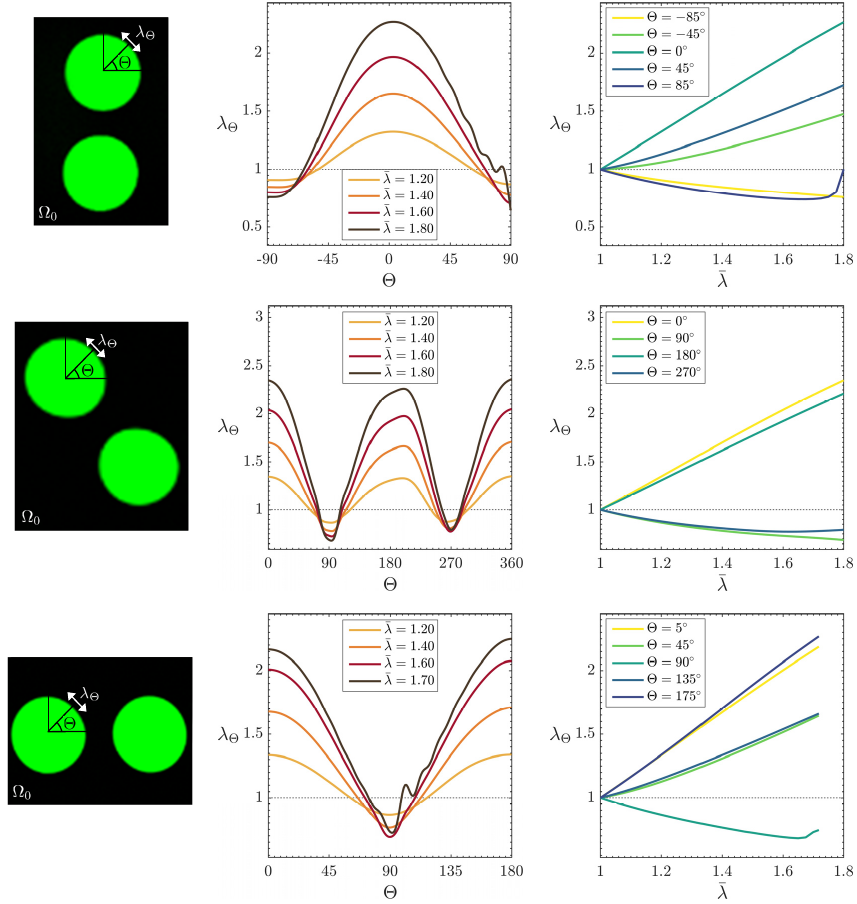}
\caption{\small The hoop stretch $\lambda_{\Theta}$ along the elastomer/inclusion interfaces of the individual inclusions in the simulations shown in Fig. \ref{Fig12} for the pairs of inclusions oriented at $0^\circ$, $45^\circ$, $90^\circ$ with respect to the loading direction. The results are shown for several values of the applied macroscopic stretch $\ol$, as a function of the hoop angle $\Theta$, and for several values of $\Theta$, as a function of $\ol$.}\label{Fig14}
\end{figure}

To further quantify the effect on the deformation of liquid inclusions from the presence of closely interacting inclusions, Fig. \ref{Fig14} presents plots for the hoop stretch $\lambda_\Theta$ along the elastomer/inclusions interfaces obtained from the simulations in Fig. \ref{Fig12} for the three pairs of inclusions oriented at $0^\circ$, $45^\circ$, $90^\circ$ with respect to the loading direction. The results are shown for several values of the applied macroscopic stretch $\ol$, as a function of the hoop angle $\Theta$, and for several values of $\Theta$, as a function of $\ol$. An immediate observation from Fig. \ref{Fig14} is that, in comparison with the results shown for an isolated inclusion in Fig. \ref{Fig9}, the hoop stretch $\lambda_\Theta$ exhibits a more asymmetric dependence on $\Theta$ and $\ol$. For all three different orientations, as for an isolated inclusion, at any given macroscopic stretch $\ol$, the largest hoop stretch is always attained at one or both ends of the equator, when $\Theta=0$ and/or $\pi$. As $\Theta$ deviates from $0$ and $\pi$ and the poles $\Theta=\pi/2, 3\pi/2$ are approached, the hoop stretch eventually changes to be compressive in a manner that is strongly dependent on the orientation of the pair of inclusions. Because of the lack of symmetry, the maximum compressive hoop stretch is not necessarily attained at the poles of the inclusions. Similar to the case of isolated inclusions, the neighborhoods of maximum compressive hoop stretch may develop creases. 

\section{Summary and final comments}\label{Sec:Final comments}

The deformation of isolated and pairs of liquid glycerol inclusions embedded in a soft PDMS elastomer was investigated by means of experiments and simulations. The focus was on large elastic deformations of initially spherical inclusions within the limit regime when elasto-capillary effects are negligible. Experiments corroborated that the deformation of the inclusions was indeed fully repeatable, reversible, non-hysteretic (i.e., elastic), as well as independent of the size of the inclusions (i.e., independent of elasto-capillary effects).

The use of fluorescent confocal microscopy allowed to directly measure the local deformation within a spatial resolution of $1$ $\mu$m and establish that the local deformation of the isolated inclusions was non-uniform and larger than that in the surrounding PDMS elastomer. Interestingly, it was also found that the large compressive stretches that develop at the poles of the inclusions when the specimens are macroscopically stretched in uniaxial tension may result in the development of creases. As expected, the deformation of the pairs of inclusions exhibited a larger non-uniformity than that found in isolated inclusions, as well as a strong dependence on the orientation of the pair with respect to the applied macroscopic stretch. 

\begin{figure}[H]
\centering
\centering\includegraphics[width=0.85\linewidth]{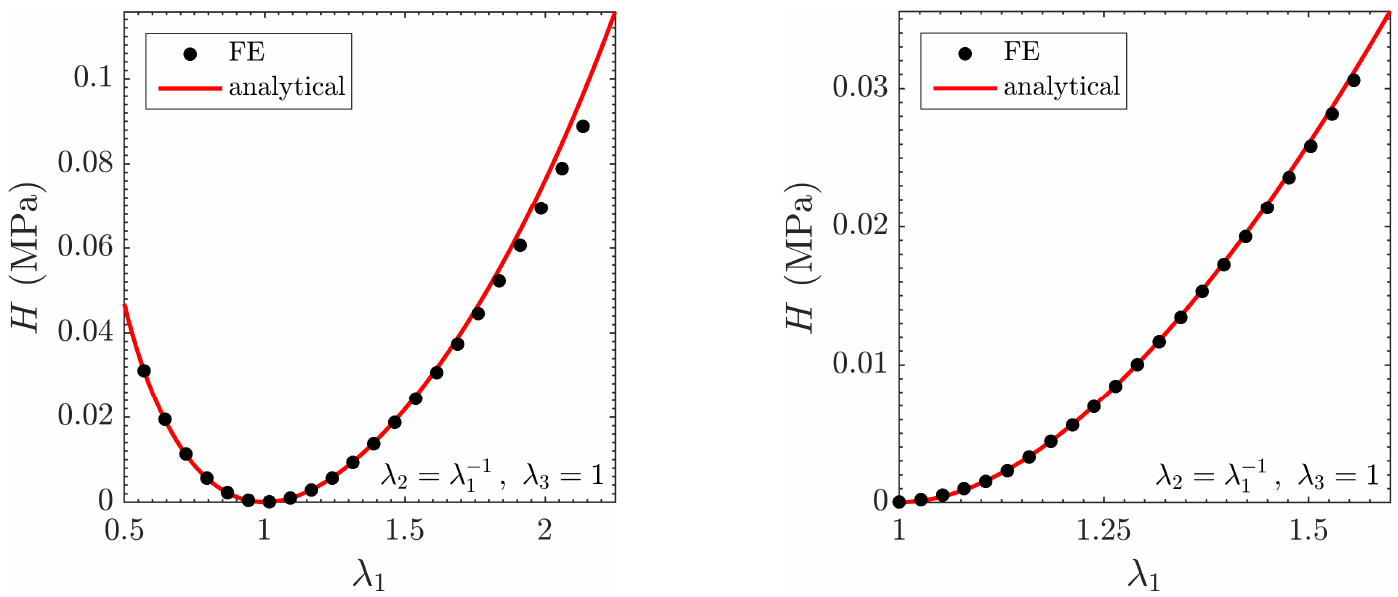}
\caption{\small The correction function $H(\bfF)$ in the effective stored-energy function (\ref{W-Dilute-Approx})$_1$ that describes the elastic response of a dilute suspension of initially spherical liquid glycerol inclusions in a soft PDMS elastomer. Part (a) shows the comparison between FE results and the approximate analytical formula (\ref{H-Approx}) for the case of axisymmetric shear when $\lambda_3=\lambda_2=\lambda_1^{-1/2}$, while part (b) shows the corresponding comparison for the case of pure shear when $\lambda_2=\lambda_1^{-1}$ and $\lambda_3=1$.}\label{Fig15}
\end{figure}
The simulations were shown to provide a highly accurate description of the deformation of the inclusions in all the specimens that were tested. This accuracy indicates that the results from such simulations can be confidently used for applications, for instance, in the construction of homogenization-based models that describe the nonlinear elastic response of suspensions of liquid inclusions in elastomers. To see this --- based on this type of simulations for a suspension of liquid inclusions in a Neo-Hookean elastomer \citep{LFLP22} and the use of a nonlinear comparison medium method\footnote{Starting with the pioneering works by \cite{TalbotWillis85} and \cite{PPC91},  linear \citep{Willis91,TalbotWillis92,Suquet93,PPC02} and nonlinear \citep{TalbotWillis94,deBotton10,LLP17b,LDLP17} comparison-medium methods have repeatedly proven extremely powerful to construct approximations for the homogenized response of the nonlinear mechanical and physical properties of composite materials from auxiliary homogenization solutions.}   \citep{LPGD13b} --- we recall that \cite{LLP17a,LLP17b} worked out the analytical result
\begin{equation*}
\overline{W}(\bfF)=\left\{\hspace{-0.1cm}\begin{array}{ll}(1-c)\Psi_{\texttt{m}}(\mathcal{I}_1), & J=1 \vspace{0.2cm}\\
+\infty, & {\rm else}\end{array}\right. \quad{\rm with}\quad\mathcal{I}_1=(1-c)^{2/3}\left(I_1-3\right)+3
\end{equation*}
for the effective stored-energy function that describes the homogenized nonlinear elastic response of an isotropic suspension of initially spherical liquid inclusions, at initial volume fraction $c$, in an incompressible isotropic elastomer with stored-energy function (\ref{LP-model})$_1$. In the dilute limit of inclusions, as $c\searrow 0$, this result reduces to
\begin{equation}\label{W-Dilute-Approx}
\overline{W}(\bfF)=\left\{\hspace{-0.1cm}\begin{array}{ll}\Psi_{\texttt{m}}(I_1)-H(\bfF)\,c+O(c^2), & J=1 \vspace{0.2cm}\\
+\infty, & {\rm else}\end{array}\right. \quad{\rm with}\quad H(\bfF)=\Psi_{\texttt{m}}(I_1)+\dfrac{2}{3}(I_1-3)\dfrac{{\rm d}\Psi_{\texttt{m}}}{{\rm d}I_1}(I_1).
\end{equation}
In turn, when applied to the stored-energy function (\ref{LP-model}) used to describe the nonlinear elastic response of the PDMS elastomer in the experiments presented in this work, the correction function in (\ref{W-Dilute-Approx}) specializes to
\begin{equation}\label{H-Approx}
H(\bfF)=\displaystyle\sum_{r=1}^2\dfrac{3^{1-\alpha_r}}{2\alpha_r}\mu_r\left(I^{\alpha_r}_1-3^{\alpha_r}\right)+
\displaystyle\sum_{r=1}^23^{-\alpha_r}\mu_r\left(I_1-3\right)I_1^{\alpha_r-1}
\end{equation}
with the material constants listed in Table \ref{Table1}. Figure \ref{Fig15} presents comparisons between the analytical approximation (\ref{H-Approx}) and FE results --- computed from the relation $H^{{\rm FE}}(\bfF)=(\overline{W}^{{\rm FE}}(\bfF)-\Psi_{\texttt{m}}(I_1))/c$ in terms of the effective stored-energy function $\overline{W}^{{\rm FE}}(\bfF)$ generated via FE --- for the cases of axisymmetric and pure shear when $I_1=\lambda_1^2+2\lambda_1^{-1}$ and $I_1=\lambda_1^2+\lambda_1^{-2}+1$, respectively. A quick glance suffices to recognize that the two results are in good agreement. 

An obvious next step is to carry out an analogous combined experimental and theoretical investigation of the deformation of isolated and pairs of liquid inclusions embedded in soft elastomers when elasto-capillary effects are dominant. Such an investigation should provide invaluable quantitative insight into the expected --- yet barely understood --- nonlinearities of the surface tension at the interfaces between liquids and highly deformable solids.

\section*{Acknowledgements}

Support for this work by the National Science Foundation through the Grants DMREF--1922371 and DMREF--1921969 is gratefully acknowledged. The confocal microscopy studies were carried out in the Carl R. Woese Institute for Genomic Biology (IGB) at the University of Illinois Urbana-Champaign.

\bibliographystyle{elsarticle-harv}
\bibliography{References}

\end{document}